\documentclass[journal abbreviation]{copernicus}

\usepackage{graphicx}

\begin{document}

\title{Signatures of Interchange Reconnection:  STEREO, ACE and Hinode Observations Combined}

\author[1]{D. Baker}
\author[2,3]{A. P. Rouillard}
\author[1,4,5]{L. van Driel-Gesztelyi}
\author[4]{P. D\'emoulin}
\author[1]{L. K. Harra}
\author[6,7]{B. Lavraud}
\author[3]{J.A. Davies}
\author[6,7]{A. Opitz}
\author[8]{J. G. Luhmann}
\author[6,7]{J.-A. Sauvaud}
\author[9]{A. B. Galvin}

\affil[1]{University College London, Mullard Space Science Laboratory, Holmbury St. Mary, Dorking, Surrey, RH5 6NT, U.K.}
\affil[2]{Space Environment Physics Group, School of Physics and Astronomy, University of Southampton, UK.}
\affil[3]{Space Science and Technology Department, Rutherford Appleton Laboratory, UK.}
\affil[4]{Observatoire de Paris, LESIA, FRE 2461(CNRS), F-92195 Meudon Principal Cedex, France.}
\affil[5]{Konkoly Observatory of Hungarian Academy of Sciences, Budapest, Hungary.}
\affil[6]{Centre d'Etude Spatiale des Rayonnements, UniversitŽ de Toulouse (UPS), France.}
\affil[7]{Centre National de la Recherche Scientifique, UMR 5187, Toulouse, France.}
\affil[8]{Space Sciences Laboratory, University of California, Berkeley, CA 94720, USA.}
\affil[9]{Institute for the Study of Earth Oceans and Space, University of New Hampshire, Durham, NH 03824, USA.}


\runningtitle{Signatures of Interchange Reconnection}

\runningauthor{Baker et al.}

\correspondence{D. Baker\\ (db2@mssl.ucl.ac.uk)}

\received{June 2009}
\pubdiscuss{} 
\revised{}
\accepted{}
\published{}


\firstpage{1}

\maketitle

\begin{abstract}
Combining STEREO, ACE and \emph{Hinode} observations has presented an opportunity to follow a filament eruption and coronal mass ejection (CME) on the 17th of October 2007 from an active region (AR) inside a coronal hole (CH) into the heliosphere.  This particular combination of `open' and closed magnetic topologies provides an ideal scenario for interchange reconnection to take place.  With \emph{Hinode} and STEREO data we were able to identify the emergence time and type of structure seen in the \emph{in-situ} data four days later.  On the 21st, ACE observed \emph{in-situ} the passage of an ICME with `open'  magnetic topology.  The magnetic field configuration of the source, a mature AR located inside an equatorial CH, has important implications for the solar and interplanetary signatures of the eruption.  We interpret the formation of an `anemone' structure of the erupting AR and the passage \emph{in-situ} of the ICME being disconnected at one leg, as manifested by uni-directional suprathermal electron flux in the ICME, to be a direct result of interchange reconnection between closed loops of the CME originating from the AR and `open' field lines of the surrounding CH.  
\end{abstract}

\keywords{Reconnection, CME, ICME, particle abundances, X-ray, EUV, \emph{in-situ} observations, solar wind}

\introduction
Interchange reconnection was defined by \citet{crooker02} to be reconnection between closed and `open' magnetic field lines.  During the process, closed field lines are `interchanged' and `open' field lines are transported, or `jump' distances determined by the length of the reconnecting closed field lines.  \citet{gosling95} and \citet{crooker02} proposed that the closed field lines in coronal mass ejections (CMEs) are `opened' or disconnected from their solar footpoints by interchange reconnection.  This occurs where `open' field lines reconnect with one `leg' of an Interplanetary CME (ICME) that is expanding into interplanetary space.  In this scenario, a large CME loop is `opened' leaving a small reconnected loop on the solar surface.  \citet{crooker02} recognized that through interchange reconnection, the total magnetic flux balance in the heliosphere is maintained as CMEs become magnetically `open', thus avoiding the `magnetic field magnitude catastrophe' \citep{gosling75}. 

Interchange reconnection may take place in any number of solar, magnetospheric, and heliospheric contexts where `open' and closed field lines exist in close proximity, e.g. at coronal hole (CH) boundaries \citep{wang93,fisk99,wang04}, between emerging flux and a nearby CH \citep{baker07}, an expanding CME structure and a nearby CH \citep{attrill06,crooker06,harra07}, in the legs of CMEs \citep{crooker02,owens07}, and in polar caps \citep{watanabe09}.

We consider a scenario where an active region (AR) has emerged in an equatorial CH.  The magnetic field configuration of the event has important implications for the eruption observed both on the Sun and in interplanetary space.
Interchange reconnection naturally may occur on the side of the AR where the magnetic field is oppositely aligned to the surrounding CH field.  Whatever the polarity of the CH, having an embedded bipole will always produce antiparallel magnetic orientation on one side of the bipole.  Expansion (e.g. driven by flux emergence and/or eruptions) of the closed field lines/loops will induce reconnection, leading to the formation of a `sea anemone' structure \citep{shibata94b,asai08}.  The anemone is characterized by loops that connect the AR's positive polarity (in this case) and the opposite negative polarity of the surrounding unipolar CH (see \citet{asai08} Figure 4 for a cartoon of an anemone AR from the side and top views).  As the AR reconnects with the CH field,
the location of `open' field is interchanged between the CH and the AR footpoints.  The consequences of interchange reconnection include X-ray jets and H$\alpha$ surges \citep{yokohama94}.  When a CME erupts from the AR, interchange reconnection leads to disconnection of the expanding CME at one of its footpoints.  The newly `opened' field line(s) are highly curved, or `refolded' on themselves (see Figure 1b in \citet{crooker02} and Figure 3 in \citet{demoulin07a}).

One of the key questions in Sun-Earth Connection science is:  How do the properties of interplanetary structures relate to their origins on the Sun?  The isolation of the AR in the CH in a very quiet period of solar activity provided an opportunity to make clear associations between solar and interplanetary signatures of interchange reconnection.  There have been only a few cases when interplanetary signatures of interchange reconnection have been definitively linked to their solar origins, most of which were done in similarly quiet solar periods \citep{attrill06,crooker06,harra07,rouillard09a}.  In this paper, we examine in detail the coronal observations of the  AR that emerged in a CH and the on-disk observations of an eruption originating from the AR and then follow the propagation of the eruption to 1 AU using 3 spacecraft.  We propose that the magnetic configuration of the AR in a CH is the most suitable environment for interchange reconnection to take place and as a result, one `leg' of the ICME which erupts is disconnected.

The paper is organized as follows:  In Section 2, we describe the remote sensing and \emph{in-situ} instruments used in our study.  Data reduction techniques are briefly mentioned.  In Section 3, we present a detailed analysis of solar and \emph{in-situ} observations of an eruption that occurred on 17 October 2007.  We also show that days prior to the eruption, transients were released intermittently from this same AR.  In Section 4, we discuss our findings which include the solar and interplanetary evidence for interchange reconnection.  Finally,  we present our conclusions in Section 5.

\section{Instrumentation and Data Reduction}
\emph{Solar TErrestrial RElations Observatory} (STEREO) \citep{kaiser08} consists of twin spacecraft, one trails (referred to as STEREO-B) the Earth while the other leads (referred to as STEREO-A).  Each STEREO spacecraft is equipped with remote sensing and \emph{in-situ} particles and field instruments for the main purpose of understanding CME initiation processes on the solar disk and to follow the propagation of CMEs into the heliosphere.  The spacecraft, launched on 26 October 2006, orbit the Sun in the ecliptic plane at a heliocentric distance close to 1 Astromomical Unit (AU).   The angle of separation between each spacecraft and the Earth increases by $22.5^{\circ}$ per year \citep{kaiser08}.  Each spacecraft carries a suite of imagers - the Sun-Earth Connection Coronal and Heliospheric Investigation (SECCHI) package \citep{howard08}.  SECCHI consists of an extreme ultraviolet imager (EUVI), two coronagraphs (COR-1 and COR-2), and the Heliospheric Imager (HI).  The EUVI observes the chromosphere and low corona in EUV emission lines at 171, 195, 284, and 304 \AA.  The HI instrument on each STEREO spacecraft comprises two wide-field, visible-light imagers, HI-1 and HI-2 \citep{eyles09,harrison08,brown09}.  The HI detectors are charge-coupled devices (CCDs) with 2048$\times$2048 pixels and nominal cadence of 40 min for HI-1 and two hours for HI-2.  1024$\times$1024 pixel synoptic science images are routinely downloaded.  HI-1 has a $20^{\circ}$ square field of view (FOV), centered at $14^{\circ}$ elongation.  The $70^{\circ}$$\times$$70^{\circ}$ FOV of the outermost HI-2 camera is centered at 53.7$^{\circ}$ elongation.  Note that the elongation of a target is defined as the angle between the observer-Sun vector and the observer-target vector. 

\begin{figure}[t]
\vspace*{2mm}
\begin{center}
\includegraphics[width=6.5cm]{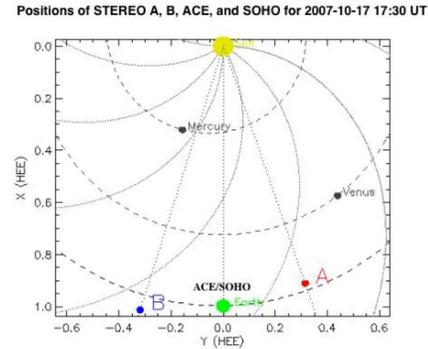}
\end{center}
\caption{Positions of the STEREO-Ahead (red) and Behind (blue) spacecraft relative to the Sun (yellow) and Earth (green).  ACE and SOHO are located at L$_{1}$.  \emph{Hinode} is in a Sun-synchronous orbit around the Earth (not shown).  Estimated Parker Spirals are shown.  (Adapted from the image produced by the STEREO Orbit Tool available at \url{http://stereo.gsfc.nasa.gov/where.shtml})}
\label{orbits}
\end{figure}

In addition to the SECCHI imaging suite described above, each of the STEREO spacecraft also carries a comprehensive suite of \emph{in-situ} instrumentation, including the PLAsma and SupraThermal Ion Composition (PLASTIC; \citep{galvin08}) and the In-Situ Measurements of Particles and CME Transients (IMPACT; \citep{luhmann08}) packages.  Magnetic field measurements from the magnetometer (MAG; \citep{acuna08}) and suprathermal electron observations from the Solar Wind Electron Analyser (SWEA; \citep{sauv08}), two components of the IMPACT package, together with the solar wind ion moments derived from measurements made by the PLASTIC package, are used in our analysis of the heliospheric consequences of the eruption.  \emph{In-situ} measurements of near-Earth solar wind electrons and ions as well as suprathermal electrons made by the Solar Wind Electron, Proton, Alpha Monitor investigation (SWEPAM; \citep{mccomas98}), solar wind composition measured by the  Solar Wind Ion Composition Spectrometer and the Solar Wind Ion Mass Spectrometer (SWICS/SWIMS; \citep{gloeckler98}) and measurements of the magnetic field by the magnetic field investigation (MAG; \citep{smith98}) onboard the Advanced Composition Explorer (ACE; \citep{stone98}) are also used in the present paper.  The ACE IMF and solar wind ion parameters are  64-second averages and the solar wind composition data are hourly averages.  The STEREO IMF and solar wind ion parameters are all 10 minute averages.

Two other spacecraft were used to complement STEREO EUVI 171, 195 and 284 \AA~on-disk observations.  \emph{Hinode's} X-Ray Telescope (XRT) \citep{golub07} is a high-resolution grazing-incidence telescope with a wide temperature coverage.  We used the thin aluminum-on-mesh  filter to observe the AR and CH evolution in the corona at a cadence of $\sim$1$\frac{1}{2}$ minutes at many time intervals around the eruption.   We examined the evolution of the photospheric magnetic field and determined AR and CH polarities using full-disk magnetograms with a 96 min cadence and a pixel size of 1.98'' that were obtained with \emph{SOlar and Heliospheric Observatory's} (SOHO's) Michelson Doppler Imager (MDI; \citep{scherrer95}).  The MDI data were corrected for the underestimation of MDI flux as discussed in \citet{green03}.  All solar data were calibrated and instrumental effects corrected for using standard SolarSoft routines for the respective instruments.

\section{Observations} 

We have used remote and \emph{in-situ} data from five spacecraft in different orbits in the following analysis.  In order to minimize any possible confusion, please refer to Figure~\ref{orbits} for the positions of STEREO-A, STEREO-B and ACE on 17 October 2007.  At the time of the observations discussed in this section, the separation angle between STEREO-A and B was $36.6^{\circ}$ and ACE and SOHO were, as always, at Earth's L$_{1}$ point.  \emph{Hinode} is in a Sun-synchronous orbit around the Earth.  Due to the different viewing angles of each spacecraft, there are offsets in X and Y solar coordinates in STEREO EUVI, SOHO, and \emph{Hinode} solar images.

\subsection{Solar On-Disk Observations}

\begin{figure}[t]
\vspace*{2mm}
\begin{center}
\includegraphics[width=7.0cm]{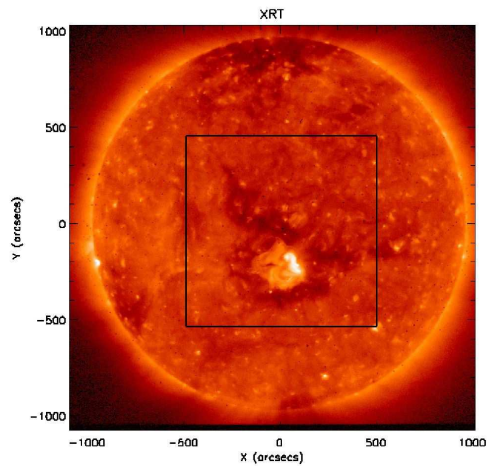}
\end{center}
\caption{\emph{Hinode} XRT thin aluminum-on-mesh filter full-disk image of the Sun on 17 October 2007 at 17:48 UT.  The AR and surrounding CH are located at the central meridian just south of the solar equator.}
\label{xrt}
\end{figure}

\begin{figure}[t]
\vspace*{2mm}
\begin{center}
\includegraphics[width=6.5cm]{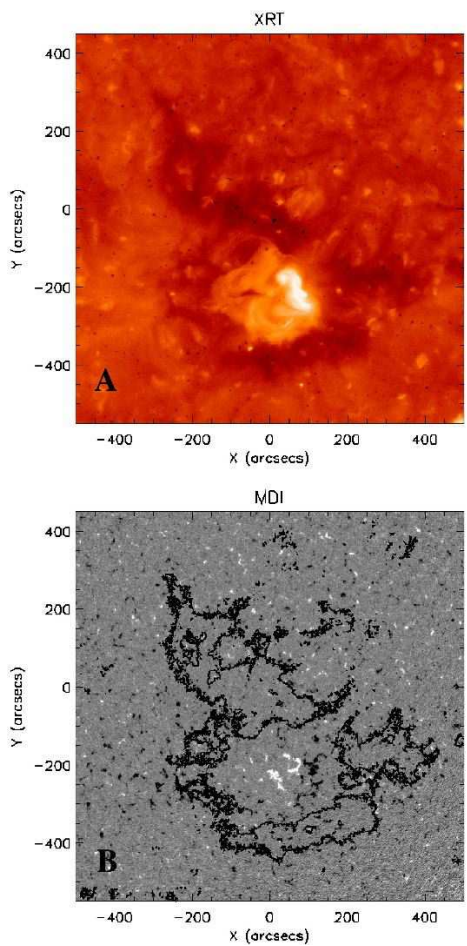}
\end{center}
\caption{A - Zoomed \emph{Hinode} XRT image contained in the black box in Figure~\ref{xrt} showing the the AR embedded in the equatorial CH.  B - Zoomed MDI magnetogram (at 17:39 UT on 17 October) with CH contour overlaid.  CH contour level is set to lie halfway between the intensity of an area of  the northern polar CH and an area of quiet Sun north of the AR \citep{attrill06}.  The AR's positive polarity (white) is to the solar east and its negative polarity (black) is to the solar west.  The CH polarity is negative.}
\label{mdi}
\end{figure}

Using multi-wavelength data, we were able to observe the evolution of the CME source AR during its passage across the solar disk.  It appeared from the eastern limb on 11 October 2007 and was located inside a low-latitude CH approximately 200'' south of the solar equator.  It crossed the central meridian at $\sim$17:00 UT on 17 October.  Images taken with \emph{Hinode's} XRT instrument provide context for the AR and the surrounding CH.  Figure~\ref{xrt} is a full disk X-ray image taken with the thin aluminum-on-mesh filter.  At this stage, the extent of the AR is approximately 300''$\times$250''.  Figure~\ref{mdi}A shows a zoomed XRT image of the field of view (FOV) contained in the black box in Figure~\ref{xrt}.  The AR structure appears to have the shape of a `sea anemone' described in the Section 1.  
 
From SOHO MDI full-disk magnetograms  the AR's signed magnetic flux was measured using the method of \citet{baker07} to be approximately 3$\times$10$^{21}$ Mx as it crossed the solar central meridian.  On 14 October, magnetograms show the following AR polarity (positive) has started to break up and disperse while the leading polarity (negative) remains essentially concentrated.  By early on the 15th, the negative field has fragmented as well.  It is clear that the AR is in the decay phase of its evolution.  Figure~\ref{mdi}B shows a zoomed MDI magnetogram timed at 17:39 UT depicting the extent of the magnetic field dispersion on the 17$^{th}$.  Contours outlining the CH as determined by the method discussed in \citet{attrill06} have been overlaid on the magnetogram.  The CH's magnetic polarity is negative and has a typical magnetic field strength of 19 G.

To the northwest, the brightest X-ray loops over the main inversion line of the AR have formed a reverse S-shaped sigmoidal structure indicating left-handed or negative helicity - unusual for an AR in the southern hemisphere.  However, we note that this CH/AR complex is north of the heliospheric current sheet (HCS) and the polarity of the CH is the same as the north polar CH \citep{simunac09}.  Sigmoidal helicity, CH polarity and the AR/CH complex position relative to the HCS should place it more in the northern hemisphere.

\begin{figure}[t]
\vspace*{2mm}
\begin{center}
\includegraphics[width=6.5cm]{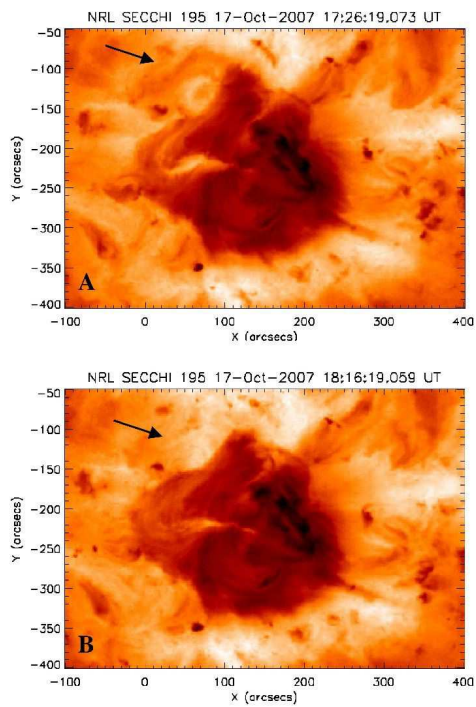}
\end{center}
\caption{STEREO-A EUVI 195 \AA~reverse color images showing before and after (17:26 UT and 18:16 UT, respectively on 17 October) the eruption to the northeast of the AR. }
\label{195b}
\end{figure}

We have used a combination of STEREO EUVI 195 \AA~and 171 \AA~data to identify the timing and location of an eruption from the AR on the 17$^{th}$.  The early eruption phase began at $\sim$17:30 UT on the north eastern (NE) side of the AR.  Figure~\ref{195b}A shows the AR at 195 \AA~prior to the eruption at 17:26 UT.  The soon-to-erupt AR loops are designated by the black arrow.  The AR loops start to expand and rise until they eventually disappear by 18:16 UT (Figure~\ref{195b}B).  A small dimming region (the bright feature in the reverse color image) is visible where the loops used to be prior to the eruption.  

\begin{figure}[t]
\vspace*{2mm}
\begin{center}
\includegraphics[width=6.5cm]{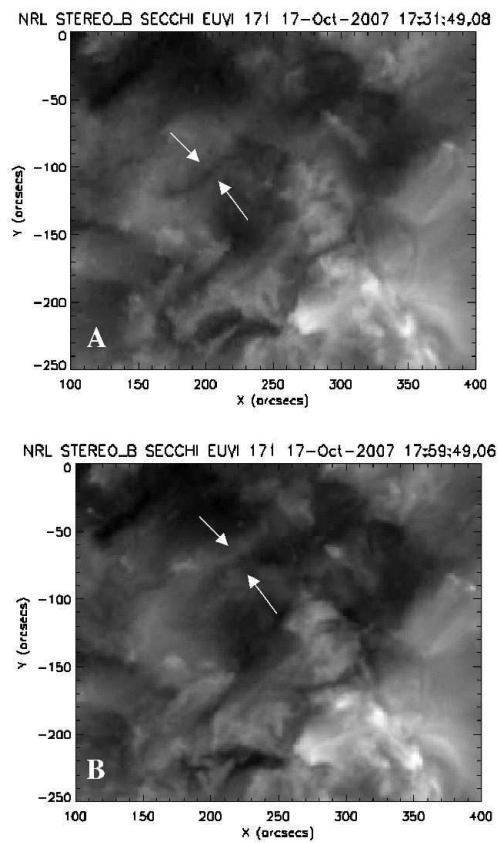}
\end{center}
\caption{STEREO-B EUVI 171 \AA~images at 17:31 (A) and 17:59 UT (B) on 17 October showing a filament (indicated by the white arrows).  The filament has started to erupt by the time of the image in B.}
\label{filament}
\end{figure}

Unlike the STEREO-A EUVI 195 \AA~images of the eruption in Figure~\ref{195b}, STEREO-B EUVI 171 \AA~images show a distinct filament in absorption prior to the eruption.  The filament lies along the polarity inversion line (PIL) where opposite polarity flux converges from early on the 17$^{th}$.  Figure~\ref{filament} shows two zoomed STEREO-B 171 \AA~images containing the filament (indicated by white arrows) located in the NE quadrant of the AR.  The eastern portion of the filament moves northwards as it rises and is noticeably displaced in the image at 17:59 UT (Figure~\ref{195b}A) relative to its position at 17:31 UT (Figure~\ref{195b}B).  (Compare the position of the eastern most segment of the filament with the Y = -100 tick mark in each image).  Movies of both STEREO-A and B EUVI 195 \AA~observations suggest the overlying loops at the location of the filament disappear by 18:30 UT which is consistent with the fact that the filament is no longer visible in the 171 \AA~images by 18:29 UT.   

\begin{figure}[t]
\vspace*{2mm}
\begin{center}
\includegraphics[width=9cm]{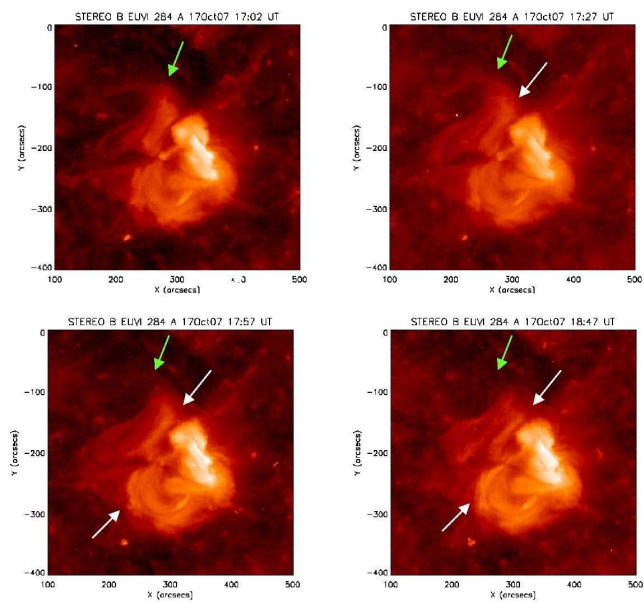}
\end{center}
\caption{STEREO-B EUVI 284 \AA~images from 17:02 to 18:47 UT on 17 October (left to right, top to bottom).  White arrows indicate anemone evolution before, during, and after the eruption.  Green arrows mark the eruption observed in EUVI 195 \AA~at $\sim$17:30 UT (cf. Figure~\ref{195b}).  The southern loops of the anemone have brightened as a result of reconnection after the eruption (images timed at 17:57 and 18:47 UT).}
\label{284images}
\end{figure}

The erupting filament lies along the magnetic inversion line surrounding (curving around) the included positive polarity. The erupted section had a SE-NW orientation, with negative (CH) polarity on the E-NE and positive (AR) polarity on the W-SW. When overlying loops  expand with the erupting filament, a favourable field line alignment for interchange reconnection would occur on the west side of the erupting loops, where  positive (anit-sunward) field lines of the expanding CME loops would meet negative open field lines of the CH.  Reconnection with CH field lines leads to the formation of new bright loops in the anemone structure connecting the positive polarity to negative fields in the NW and SW. These brightened loops are best visible in STEREO-A 284 \AA~images (Figure~\ref{284images}). 

\subsection{STEREO HI Observations}
\begin{figure}[t]
\vspace*{2mm}
\begin{center}
\includegraphics[width=6.0cm]{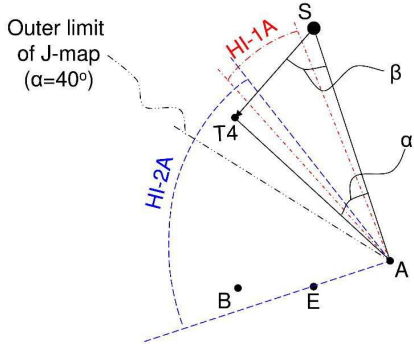}
\end{center}
\caption{View of the ecliptic from solar north with the position of the Sun (S), Earth (E), STEREO-A (A), and STEREO-B (B) spacecraft on 17 October 2007.  The limits of the fields of view of HI-1A and HI-2A imagers in the ecliptic plane are shown in red and blue, respectively.  The trajectory of transient T4 is shown by a black arrow, the angles $\alpha$ and $\beta$ which define the position of T4 in the ecliptic plane uniquely are also shown.  The outer limit of the J-maps of Figure~\ref{jmap1} ($40^{\circ}$)  is also shown.}
\label{hifov}
\end{figure}

\begin{figure}[t]
\vspace*{2mm}
\begin{center}
\includegraphics[width=6.5cm]{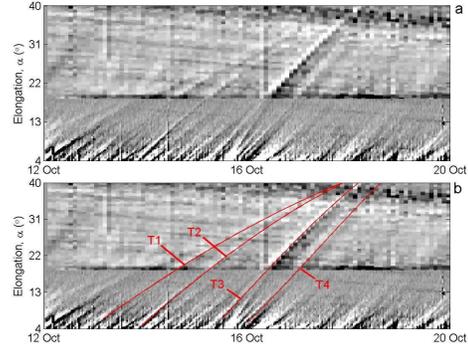}
\end{center}
\caption{Two identical J-maps constructed along PA=100$^{\circ}$ using HI-1/2A running difference images.  In  b the results of fitting the tracks (T1 to T4) are shown as red lines superposed on the J-map.} 
\label{jmap1}
\end{figure}

Figure~\ref{hifov} presents a view of the ecliptic plane from solar north on 17 October 2007 showing the relative positions of the Sun (S), STEREO-A (A), STEREO-B (B) and the Earth (E).  The elongation (see Section 2) extents in the ecliptic plane of the FOVs of HI-1 and HI-2 on STEREO-A (termed HI-1A and HI-2A) are marked in red and blue, respectively.  Figure~\ref{jmap1}a presents a time-elongation map (or J-map) derived by plotting strips from HI-1A and HI-2A running difference images, extracted along the Position Angle (PA) of $100^{\circ}$ which intersects the latitude of the AR when crossing the plane of the sky as viewed from STEREO-A, as a function of elongation (along the ordinate) and time (along the absissa).  The use of difference images minimizes the contribution of the stable F-corona and is a very useful technique for highlighting faint propagating features.  The J-map shown in Figure~\ref{jmap1} covers the interval from 12 October to 20 October 2007 and is based on the technique reported by \citet{davies09}.  The J-maps are truncated to $40^{\circ}$ elongation due to the presence of the Milky Way in the outer edge of the HI-2A field of view.  Many structures erupted during this interval, in particular tracks appear to converge suggesting the passage of a corotating solar source of transients.  \citet{rouillard08} and \citet{sheeley08a,sheeley08b} showed that the apparent acceleration/deceleration at the elongations covered by HI J-maps is mostly an effect of projection geometry.  The elongation, $\alpha$, labeled in Figure~\ref{hifov}, of a point T (e.g. T4) in the solar wind and observed by the STEREO-A spacecraft, is defined as the Sun - STEREO-A - T angle, being zero at Sun center.  The angular separation between the Sun - STEREO-A spacecraft line and the direction of propagation of the point (labeled $\beta$ in Figure~\ref{hifov}) equates to the longitude separation in an ecliptic-based heliocentric coordinate system when the transient propagates in the ecliptic plane. The elongation variation, $\alpha$(t), of a solar wind transient depends upon its radial speed, Vr, and the angle  $\beta$ \citep{rouillard08,rouillard09a}.  Best-fit values of these parameters can, therefore, be extracted from the elongation variation recorded by HI.  Each clearly traceable track was fitted using the technique; the red lines superposed on the same J-map as Figure~\ref{jmap1}a and shown in Figure~\ref{jmap1}b correspond to the best-fit line of the apparent speed variation.  The predicted speed, Vr (km $s^{-1}$), longitude separation $\beta$ ($^{\circ}$), elongation $\alpha$$_{\circ}$ ($^{\circ}$),  and inferred coronal height, D$_{\circ}$, at which the transient track is first fitted, and the estimated launch date/time of transients T1 to T4 are listed in Table~\ref{table1}. 

\begin{table}[t]
\caption{The predicted speed, Vr (km s$^{-1}$), longitude separation, $\beta$ ($^{\circ}$), elongation, $\alpha$$_{\circ}$ ($^{\circ}$), and inferred coronal height, D$_{\circ}$ at which the transient track is first fitted and the estimated launch date/time of transients T1 to T4.}
\vskip4mm
\centering
\begin{tabular}{ccccc}
\tophline
 Transient& T1 & T2 & T3 & T4\\
\middlehline
Vr (km s$^{-1}$)&260$\pm$20 &320$\pm$20&385$\pm$32&363$\pm$27\\
$\beta$ ($^{\circ}$)&90$\pm$8&90$\pm$5&79$\pm$10&63$\pm$11\\
$\alpha$$_{\circ}$ ($^{\circ}$) &6.46&4.68&6.43&5.84\\
D$_{\circ}$ (AU) &0.1095&0.0791&0.1086&0.1035\\
\middlehline
Launch Time: &&&&\\
yy/mm/dd&07/10/12&07/10/13&07/10/15&07/10/15\\
hh:mm &10:52 UT &11:58 UT &00:38 UT &13:12 UT\\
\bottomhline
\end{tabular}
\label{table1}
\end{table}

As expected from the convergence of tracks and a corotating source region, the longitude separation between the source region and STEREO-A is decreasing with time (i.e. converging tracks). We determined the location of the source-region in the lower corona of each transient by using a ballistic back-mapping of the transient position assuming constant speed (i.e. ignoring acceleration effects) and a solar rotation period as seen from STEREO-A of 28.4 days. The solar rotation period as seen from STEREO-A is longer than the well-known 27.27 day period as viewed from Earth because STEREO-A is propagating faster than the Earth around the Sun. The ballistic back-mapping assumes radial propagation of the transient and has its limitations.  Our estimate of the source region is accurate only to first approximation.

\begin{figure}[t]
\vspace*{2mm}
\begin{center}
\includegraphics[width=6.0cm]{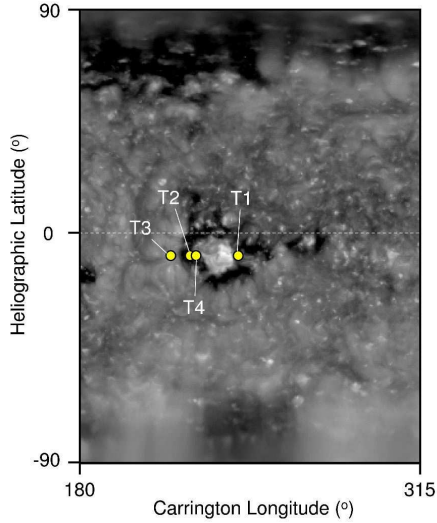}
\end{center}
\caption{The subset of a Carrington map (latitude versus longitude map for CR 2062) created from central meridian EUV observations at 195 \AA~made by STEREO-A.  The small yellow disks mark the estimated launch-site of each transient tracked in HI using the calculated kinematic properties listed in Table~\ref{table1}.}
\label{jmap2}
\end{figure}

The launch-sites of T1 to T4 are shown in Figure~\ref{jmap2} as yellow disks on a subset of a Carrington map (latitude versus longitude map for CR 2062) created from central meridian EUV observations at 195 \AA~made by STEREO-A.  Three of the transients emerge in the vicinity of the AR (T1, T2 and T4). They appear to emerge from the boundary of the AR with the CH although the errors  of the estimated longitude of propagation are around $8^{\circ}$  (average of the mean error in angle $\beta$ in Table~\ref{table1}) and while we can tell whether a transient propagated along a solar radial rooted on the eastern or western boundary of the AR, we cannot tell if the eruption occurred on the boundary or toward the center of the AR. 

The HI observations suggest that the AR is continually releasing transients in the solar wind.  We see indications of two of the transients, T2 and T4, in GOES X-ray flux curves at approximately 12:30 UT on 13 October and 12:00 UT on 15 October, respectively (\texttt{http://www.swpc.noaa.gov/Data/goes.html}).  STEREO-B EUVI 284 \AA~movies (STEREO movie maker at \texttt{http://stereo-ssc.nascom.nasa.gov}) show what appear to be transients T2, T3, and T4 erupting within a few hours of the estimated launch times on the side of the AR indicated in Figure~\ref{jmap2}.  This type of intermittent release of transients has been observed by HI during July 2007 and September 2007 and comparison of HI images with \emph{in-situ} observations during spacecraft-impacting events showed that these transients had flux-rope topologies \citep{rouillard09b,rouillard09c,rouillard09a}.  

The eruption observed in STEREO EUVI data on 17 October at $\sim$17:30 was close to Sun-center, therefore, as the transient propagates towards STEREO-A, it is not close to the Thomson surface (or surface of maximum scatter).  Furthermore, the eruption did not lift much plasma from the lower corona.  These two facts combined meant the trace was not strong enough to be seen in the J-map in Figure~\ref{jmap1}.  

\subsection{\emph{In-situ} Observations}
We established the on-disk source of the 17 October eruption and showed with HI observations that the AR embedded in a CH has produced several transients within a few days prior to the eruption.  Now we identify signatures of the eruption four days later in the solar wind data using STEREO-A, STEREO-B and ACE spacecraft.  All three spacecraft were located in or close to the ecliptic plane which intersected the solar surface $5^{\circ}$ north of the AR in the surrounding CH area at the time of the event.  Though none of the spacecraft was radially aligned with the AR, all spacecraft detected the fast flows from the surrounding CH and in particular, the trailing edge of the corotating fast stream (rarefaction region).  The eruption observed by STEREO EUVI and \emph{Hinode} XRT occurred on the NE boundary of the AR close to the latitudes of the three spacecraft.  

\subsubsection{Interplanetary ACE In-situ Observations}
\begin{figure}[t]
\vspace*{2mm}
\begin{center}
\includegraphics[width=8.0cm]{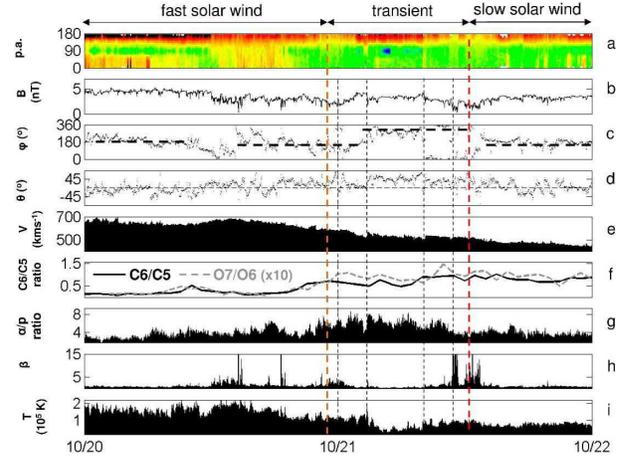}
\end{center}
\caption{In-situ data measured during the passage of the ICME.  Panel a: The 272 eV electron pitch angle [p.a.] distributions recorded by the ACE spacecraft;  Panel b: the magnetic field strength [B (nT)];  Panel c: the azimuth of the magnetic field [$\varphi$ ($^{\circ}\mathrm{}$)]; Panel d: the elevation of the magnetic field [$\theta$ ($^{\circ}\mathrm{}$)];  Panel e: the solar wind speed [V (km s$^{-1}$)];  Panel f: the charge state ratios (C6/C5) (black line) and (O7/O6) (grey dashed line) measured by SWICS onboard the ACE spacecraft;  Panel g: the alpha to proton ratio [$\alpha$/p as a percentage ($\%$)];  Panel h: solar wind plasma beta [$\beta$];  Panel i: solar wind ion temperature [T (K)].  Red dashed lines define the interval of the transient passage defined as the combined changes in magnetic field strength, alpha to proton ratio and plasma beta.  Vertical black dashed lines indicate the times of sharp discontinuities in magnetic field direction, strength, and alpha to proton ratio.}
\label{ace2}
\end{figure}

Figure~\ref{ace2} shows ACE data for the interval 20 - 22 October 2007.  On the 21$^{st}$ at 04:00 UT, 4 days after the eruption to the NE of the AR, a sharp change from $175^{\circ}$ (sunward pointing magnetic field or negative polarity footpoints) to $350^{\circ}$ (anti-sunward pointing magnetic field or positive polarity footpoints) is observed in the azimuth angle of the magnetic field direction (Figure~\ref{ace2}c).  The azimuth angle remains at $350^{\circ}$ for 10 hours and then changes back to $175^{\circ}$ (cf. black dashed line in Figure~\ref{ace2}c for guidance).  This change in azimuth angle is not associated with a reversal of the strahl as seen in the pitch angle distribution of suprathermal electrons (272eV) which remains anti-field-aligned at $175^{\circ}$ throughout the entire interval (see Figure~\ref{ace2}a).  The boundaries of this anomalous field direction are indicated by sharp increases in plasma $\beta$ (Figure~\ref{ace2}h).

The passage of the transient (see fast solar wind, transient, and slow solar wind labels at the top of Figure~\ref{ace2}) is marked by a change in solar wind composition with enhanced charge state ratios (Figure~\ref{ace2}f).  The oxygen charge state ratio of the solar wind as measured by n(O$^{7+}$)/n(O$^{6+}$) changes from $<$0.05 to $>$0.14 and the carbon charge state ratio as measured by n(C$^{6+}$)/n(C$^{5+}$) changes from 0.1 to 0.95 (oxygen ratio is the gray dashed line and carbon ratio is the black solid line in Figure~\ref{ace2}f).  The structure is also characterized by enhanced alpha to proton ratio (Figure~\ref{ace2}g) and lower solar wind temperature (Figure~\ref{ace2}i).  A preliminary analysis, not shown here, of the correlation between the components of the solar wind velocity and the magnetic field vectors reveals the presence of large amplitude Alfv\'en waves inside the transient.

A 500 km s$^{-1}$~constant speed backmapping of the arrival time of the structure observed \emph{in-situ} on 21 October suggests a launch-time on the solar surface on 17 October at 19 UT (see Section 3.2 for discussion of the method).  The eruption observed on the NE side of the AR and timed at $\sim$17:30 on 17$^{th}$ is very near the launch time estimated for the departure of the transient.

\subsubsection{Interplanetary STEREO-A and B In-situ Observations}
In mid-October, STEREO-A observed a trailing edge of a corotating interaction region (CIR) followed by an increase in solar wind speed forming a second CIR.  The second CIR (on 22/23 October) is likely to be associated with the part of the CH located on the eastern side of the AR.  The AR causes a dip in solar wind speed from above 600 km s$^{-1}$ to just above 500 km s$^{-1}$ between the two fast solar wind streams of the surrounding CH.  Unlike ACE observations, the STEREO-A data is dominated by pure solar wind as there is little evidence for transient structures with the exception of one small high $\beta$ structure.

\begin{figure}[t]
\vspace*{2mm}
\begin{center}
\includegraphics[width=9.0cm]{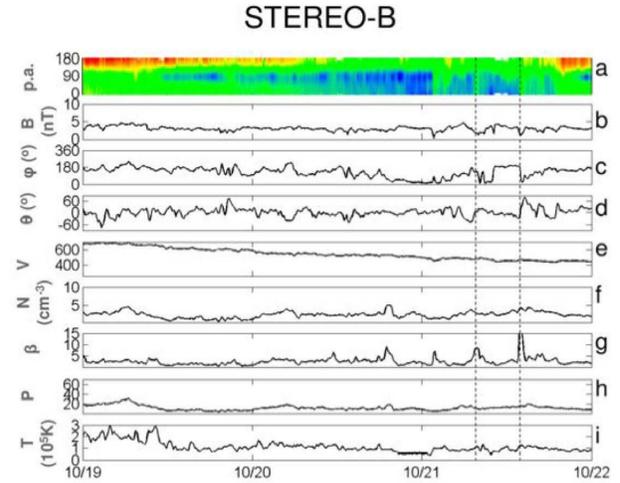}
\end{center}
\caption{STEREO-B \emph{in-situ} data.  Panel a: The 272 eV electron pitch angle [p.a.] distributions recorded by the STEREO spacecraft;  Panel b: the magnetic field strength [B (nT)];  Panel c: the azimuth of the magnetic field [$\varphi$ ($^{\circ}\mathrm{}$)]; Panel d: the elevation angle of the magnetic field [$\theta$ ($^{\circ}\mathrm{}$)];  Panel e: the solar wind speed [V (km s$^{-1}$)];  Panel f: the number density [N (cm$^{-3}$)];  Panel g: solar wind plasma beta [$\beta$];  Panel h:  pressure [nPa];  Panel i: solar wind ion temperature [T (K)].  Dashed lines mark high plasma $\beta$ structures.}
\label{stereoa}
\end{figure}

STEREO-B measured the highest speed solar wind (Figure~\ref{stereoa}e) and the trailing edge of a CIR.  There are suprathermal electrons at $0^{\circ}$ (Figure~\ref{stereoa}a) escaping from the region of high speed streams on 19 and 20 October, as was observed at ACE prior to the transient in Figure~\ref{ace2}a.  Later, on 21 October, during the transition from fast to slower solar wind there are two high plasma $\beta$ structures marked by dashed vertical lines in Figure~\ref{stereoa}.  STEREO-B is observing high-plasma $\beta$ structures with field line rotations which may be more evidence of transients but there is not enough information available to relate them to the 17 October transient.  
 
\section{Discussion}

\subsection{On-Disk Evidence of Interchange Reconnection - Presence and Evolution of Anemone Structure}

We have proposed that an AR embedded in a low-latitude CH provides the ideal scenario for interchange reconnection to take place.  Oppositely aligned magnetic field of the AR and CH will naturally exist somewhere in the magnetic configuration.  STEREO EUVI 171, 195, and 284 \AA~observations showed a filament eruption/CME took place on the NE side of the AR where its positive polarity is surrounded by the unipolar negative polarity of the CH.  There is little doubt that the magnetic configuration is ideal for interchange reconnection, but what observational evidence, if any, confirms that reconnection did in fact take place?  One possible answer lies in the anemone structure observed in \emph{Hinode's} XRT and STEREO EUVI 284 \AA~images.  \citet{crooker06} interpreted bright X-ray regions or anemones, associated with CME source regions, to be the X-ray signatures of interchange reconnection.  Like \citet{crooker06}, we interpret the anemone to be the result of interchange reconnection induced by several episodes of eruption-driven expansion of the AR.  This is consistent with previous studies of anemone ARs that show such structures sometimes generate filament eruptions \citep{chertok02} and CMEs \citep{asai09}.  

\subsection{ICME Signatures and Characteristics}
The link between transient structures such as CMEs and their interplanetary manifestations, ICMEs, is not direct.  Transients interact with the ambient solar wind so that by the time \emph{in-situ} instruments measure their plasma properties at 1 AU, the transients have been modified to some extent.  The link between CMEs and ICMEs is weakened further by limitations posed by spacecraft position relative to ICME passage.  In spite of these concerns, different signatures are used to identify ICMEs \emph{in-situ}, some of which we employ here.  (See \citet{wimmer06b} for a review of ICME signatures).

Analysis of \emph{in-situ} data covering the period from 19 October to 22 October revealed varying degrees of ICME signatures of the filament eruption/CME from the AR on 17 October.  Any transients went virtually undetected in STEREO-A data which were dominated by pure solar wind.  STEREO-B data revealed little more.  The fact that neither STEREO A nor B observed the transient set an upper boundary to the angular extent of any transient/ICME as the spacecraft were separated by $36.6^{\circ}$ at the time, though there is a slight possibility that the
non-detection is due to the southward deflection of the ICME.

In the ACE data, the transient passage is marked by a slight increase in magnetic field strength towards the center of the transient and elevation angles (out of ecliptic fields), enhanced alpha to proton ratio (Figure~\ref{ace2}g), lower plasma $\beta$ (Figure~\ref{ace2}h) and lower temperatures (Figure ~\ref{ace2}i), suggesting the passage of an ICME.  The rotations of the azimuth and elevation angles are not smooth and the variance of magnetic field does not drop inside the transient which suggests that either the event is not a magnetic cloud (MC) (a well-defined subset of ICMEs) by the strict definition of \citep{burlaga81} or that ACE intersected the edges of the MC only and the spacecraft failed to sample the central axis of the flux-rope.  Wilcox Solar Observatory (WSO) synoptic magnetic maps and source-surface computations (\url{http:wso.stanford.edu/MeanField}) show that the warped HCS runs south of the CH/AR complex at about S $30^{\circ}$.  Since open field of CHs are known to deflect CMEs \citep{gopalswamy09}, it is plausible that our CME was `channeled' by its surrounding CH field towards the HCS, i.e. southward. This may provide an explanation why ACE, situated in the ecliptic, observed only the flanks of the ICME.

\begin{figure}[t]
\vspace*{2mm}
\begin{center}
\includegraphics[width=8.3cm]{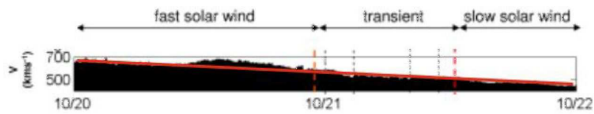}
\end{center}
\caption{Velocity profile of the transient as observed by ACE.  The near linear dependence of observed velocity with time and expansion rate $\zeta$$ \approx$0.7 are consistent with \citet{demoulin08} results for expanding ICMEs/MCs.}
\label{zeta}
\end{figure}

\citet{demoulin08} proposed a model for the expected \emph{in-situ} velocities of expanding ICMEs.  The model yields a nearly linear temporal dependence of the velocity which is consistent with observed velocity profiles of 26 MCs not overtaken by fast solar wind streams.    They conclude for most ICMEs/MCs the observed velocity profile is mainly due to expansion which can be described by the normalized expansion factor $\zeta$ computed from the slope of the velocity profile.  \citet{demoulin08} results showed $\zeta$ = 0.8$\pm$0.2 for the 26 MCs.  The ICME analyzed in this paper has a typical velocity profile (see Figure~\ref{zeta}) and expansion rate ($\zeta$ $\sim$0.7) within the range of results \citet{demoulin08} found for their sample of MCs.  

\begin{figure}[t]
\vspace*{2mm}
\begin{center}
\includegraphics[width=6.0cm]{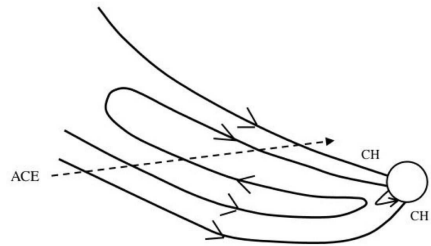}
\end{center}
\caption{Cartoon of the ICME after disconnection of the positive `leg'.  ACE trajectory is indicated by the dashed arrow.}
\label{cartoon}
\end{figure}

MCs are often enriched in alpha to proton ratio \citep{hirshberg72,neugebauer97,zurbuchen02}.
Normal solar wind alpha to proton ratio levels range from 3 to 5$\%$ \citep{neugebauer81,schwenn90}.  
The alpha to proton ratio increases sharply inside the whole ICME structure from 0.01 to 0.06, just below the 0.08 limit used to distinguish ICMEs \citep{hirshberg70,hirshberg72,neugebauer81,borrini82} which could be a result of dynamical effects lower in the corona.  

As described in Section 3.3.1, the passage of the ICME is marked by increases in oxygen and carbon charge state ratios.  The oxygen charge state ratio has a relatively fast freeze-in process in the low corona, therefore, it is considered to be a good measure of the source of the solar wind (\citet{zurbuchen02} and references therein).  In addition, \citet{henke98} found that ICMEs with magnetic cloud structure have an enhanced oxygen charge state ratio compared with non-cloud ICMEs.  The n(O7+)/n(O6+) charge-state ratio observed inside the transient is enhanced compared to fast solar wind but is much less than the n(O7+)/n(O6+)$\geq$1 sometimes observed in ICME magnetic clouds (Zurbuchen et al., 2002).  The ICME itself separates pure fast solar wind from slow solar wind and is located near the stream interface (SI). The SI on this side of the CIR is often overlooked but is a simpler boundary to study than the SI located on the compression region side of the CIR \citep{cartwright08,kilpua09}.  Thus as a counterpart to transients found to be entrained in CIRs, we have presented a clear case of an ICME in the rarefaction region. 

There is good evidence that the transient observed \emph{in-situ} 4 days after the eruption to the northeast of the AR on 17 October is a transient/ICME.  Indeed, this transient has some characteristics of a MC as suggested by the enhanced alpha to proton and oxygen charge state ratios and the level of the expansion factor $\zeta$, though as previously stated, ACE probably has not sampled the flux rope's central axis.

\subsection{In-situ Evidence of Interchange Reconnection - Disconnection of One Side of the ICME}
Suprathermal electrons can be used as an indicator of magnetic connection to the Sun.  Counterstreaming (or bidirectional) electrons are one of the benchmark indicators of the passage \emph{in-situ} of  ICMEs connected at both ends in the photosphere \citep{gosling87}.  However, not all ICMEs are associated with counterstreaming suprathermal electrons.  \citet{shodhan00} found that most MCs contained a mixture of closed and `opened' field lines at 1 AU and \citet{crooker04b} found an average of $55\%$ closed in their study of 31 MCs at 5 AU.  These results are consistent with \citet{riley04}.  The ACE suprathermal electron pitch angle spectrogram shows no indication of counterstreaming electrons during the passage of the ICME, suggesting the ICME is `open' and therefore, disconnected from the Sun on one side.  

\citet{crooker02} and \citet{owens06} proposed that interchange reconnection is the mechanism by which ICME field lines are `opened'.  In a case study of a CME on 12 May 1997, \citet{attrill06} and \citet{crooker06} independently concluded that long-lasting interchange reconnection occurred throughout the CME release process, disconnecting the negative leg of the CME.  \citet{rouillard09a} recently observed a CME transient located on the anti-sunward flank of the SI associated with the compression side of a CIR. Their analysis also revealed that the transient was only connected at one end to the photosphere inferring interchange reconnection as the mechanism by which the magnetic field lines of the transient became `open'.  

The event described here is another clear example of the process of interchange reconnection opening the ICME field.  High suprathermal electron fluxes are predominantly at $175^{\circ}$ throughout the passage indicating sunward pointing magnetic field or negative polarity footpoints.  The CH field surrounding the AR is also negative, suggesting  interchange reconnection took place between the positive field of the AR and the negative field of the CH, opening the positive polarity `leg' of the  ICME to the east and leaving the negative polarity connected to the Sun.

Another possible \emph{in-situ} signature of interchange reconnection is the observation of so-called refolded magnetic field lines (RFL) in the ICME.  The azimuth angle of the magnetic field inside the transient (Figure 9c) at first stays around $180^{\circ}$, followed by a sharp change in the azimuth angle of the magnetic field direction  to $\sim$$340^{\circ}$ that lasted for 10 hours before reverting back. This change in the field orientation was not associated with a reversal of the strahl indicating that magnetic field line changed pointing direction but did not change polarity.  \citet{crooker04a} proposed this type of event is a locally refolded magnetic field line (RFL).

Recalling our on-disk observations of the erupting filament described in Section 3.1, we can interpret the sharp changes in  azimuth angle  observed by ACE:  as shown in the cartoon in Figure~\ref{cartoon}, the expanding loops of the ICME are favourably oriented for reconnection with open CH field towards the west, disconnecting the positive magnetic footpoints of the CME/ICME.  (Note that as described in Section 3.1, the eruption occurred at the NE periphery of the AR, not at its main inversion line).  With the magnetic field orientation observed in the eruption we expect ACE to intercept open field lines of the CH then perhaps RFLs, first intercepting the negative fold of the RFL, which appears as the 2-3 hour long $\sim$$180^{\circ}$ oriented azimuth  angle first part in the transient/ICME. A sharp change is observed when ACE meets the positive leg of the expanding loop-like ICME structure. This phase lasts for about 10 hours.  After that we observe the negative leg of the ICME fieldlines which remain connected to the Sun. In this scenario the transient is more extended than indicated in Figure~\ref{ace2}, which is not contradicted by solar wind composition, temperature, nor magnetic field measurements. 

\section{Conclusions}
We have provided evidence that an eruption from the Sun on 17 October was linked to a transient observed \emph{in-situ} at 1 AU four days later.  The transient showed many of the properties of an ICME of `open' topology.  A unique magnetic configuration consisting of an AR embedded inside a CH proved to be highly favorable and effective for interchange reconnection to take place as the eruption-driven expansion of the AR induced reconnection between oppositely aligned closed positive field of the AR and CH's `open' negative field.  Two clear direct consequences of interchange reconnection were observed, one on the solar surface - the presence and evolution of an anemone coronal loop structure in the AR while a series of ejecta were traced in HI data to erupt from the AR/CH complex, and \emph{in-situ} - the disconnection of one side of the ICME perhaps accompanied by RFL topology observed by ACE.

There are still some open questions requiring further investigation.  Though we were able to identify two consequences of interchange reconnection, are there any other more subtle indications of the process?  The substructure clearly evident within the large-scale structure of the transient may provide clues to other manifestations of interchange reconnection.  The sharp increases in Alfv\'enicity noted in the ACE data that appear to mark the boundaries of different solar wind structures may suggest the possibility that distinct bundles of magnetic field lines pass over the ACE spacecraft over successive 2-3 hour period \citep{borovsky08}.  Are we observing the consequences of ongoing interchange reconnection where the ICME loops expand, becoming 'frayed' as they continually, from the low corona through their propagation to 1 AU, reconnect with the surrounding `open' (CH) field, creating distinct bundles of field lines with a highly curved (refolded) shape?  Are the Alfv\'en waves signatures of these discreet reconnection processes or were they initiated by the fast-expanding CME along the CH fieldlines?  These questions will be addressed in future work.

\begin{acknowledgements}
We would like to thank the referees and the editor for their suggestions which help to strengthen the paper.  DB thanks STFC for support via PhD studentship.  LvDG's work was partially supported by  the European Commission through the SOTERIA Network (EU FP7 Space Science Project No. 218816).   The SECCHI data used in this paper is stored at the World Data Center C1 Chilton, UK. We thank the RAL HI team for calibrating and preparing the HI data shown in this paper. The STEREO/SECCHI data are produced by a consortium of RAL (UK), NRL (USA), LMSAL (USA), GSFC (USA), MPS (Germany), CSL (Belgium), IOTA (France) and IAS (France). The PLASTIC and IMPACT data are produced by a consortium of the CESR (SWEA principal investigator.: J.-A. Sauvaud), the University of New Hampshire (STEREO PLASTIC investigation, principal investigator A.B. Galvin and NASA Contract NAS5-00132.) and the University of California (IMPACT principal investigator: J.G. Luhmann).  We thank the ACE SWEPAM, ACE MAG, ACE SWICS/SWIM instrument teams and the ACE Science Center for providing the ACE data.  \emph{Hinode} is a Japanese mission developed and launched by ISAS/JAXA, with NAOJ as domestic partner and NASA and STFC (UK) as international partners.  It is operated by these agencies in co-operation with ESA and NSC (Norway).  
 
\end{acknowledgements}

\bibliographystyle{copernicus}
\bibliography{Bibtex}  
\IfFileExists{\jobname.bbl}{} {\typeout{}
\typeout{****************************************************}
\typeout{****************************************************}
\typeout{** Please run "bibtex \jobname" to obtain} \typeout{**
the bibliography and then re-run LaTeX} \typeout{** twice to fix
the references !}
\typeout{****************************************************}
\typeout{****************************************************}
\typeout{}}

\end{document}